%% file: icse27.tex
\documentclass[conference]{IEEEtran}
\IEEEoverridecommandlockouts

\usepackage{subcaption}
\usepackage{enumitem}
\usepackage{array}
\usepackage{wrapfig}
\usepackage{amsmath,amsfonts}
\usepackage[noend]{algpseudocode}
\usepackage{graphicx}
\usepackage{textcomp}
\usepackage{float}
\usepackage{listings}
\usepackage{xspace}
\usepackage{multirow}
\usepackage{amsthm}

\usepackage{balance}
\usepackage{algorithm}
\usepackage{algpseudocode}
\usepackage{colortbl}
\usepackage{amssymb}

\usepackage[skins]{tcolorbox}
\usepackage{xcolor,pifont}
\usepackage{multicol}
\newcommand*\colourcheck[1]{%
	\expandafter\newcommand\csname #1check\endcsname{\textcolor{#1}{\ding{52}}}%
}
\colourcheck{blue}
\colourcheck{green}
\colourcheck{red}
\newtcolorbox{boxB}[2][]{%
  enhanced,colback=white,colframe=black,coltitle=black,
  sharp corners,
  toprule=1.0pt,
  rightrule=0.3pt,
  leftrule=0pt,
  bottomrule=0pt,
  fonttitle=\itshape\scshape\large,
  left=0pt,right=5pt,top=5pt,bottom=3pt,
  attach boxed title to top right={yshift=-0.3\baselineskip-0.4pt,xshift=-5mm},
  boxed title style={tile,size=minimal,left=0.2mm,right=0.5mm,
    colback=white,before upper=\strut},
  title=#2,#1
}

\newcommand{\tool}{\textsc{Tessara}\xspace}
\newcommand{\model}{\textsc{SAS}\xspace}

\def\BibTeX{{\rm B\kern-.05em{\sc i\kern-.025em b}\kern-.08em
    T\kern-.1667em\lower.7ex\hbox{E}\kern-.125emX}}

\newboolean{showcomments}
\setboolean{showcomments}{true}
\ifthenelse{\boolean{showcomments}}
 { \newcommand{\mynote}[2]{
      \fbox{\bfseries\sffamily\scriptsize#1}
        {\small$\blacktriangleright$\textsf{\emph{#2}}$\blacktriangleleft$}}}
        { \newcommand{\mynote}[2]{}}

\usepackage{tikz}

\newcolumntype{L}[1]{>{\raggedright\arraybackslash}p{#1}}
\newtheorem{Observation}{Observation}

\newcommand{\code}[1]{{\footnotesize\texttt{#1}}}
\usepackage{amsthm}
\definecolor{dkgreen}{rgb}{0,0.6,0}
\definecolor{gray}{rgb}{0.5,0.5,0.5}
\definecolor{lightgray}{rgb}{211, 211, 211}
\definecolor{mauve}{rgb}{0.58,0,0.82}

\definecolor{custom-blue2}{rgb}{0,0,0}
\definecolor{custom-red}{rgb}{0,0,0}

\definecolor{c1}{HTML}{f4cccc}
\definecolor{c2}{HTML}{f5cdcd}
\definecolor{c3}{HTML}{fffcfc}
\definecolor{c4}{HTML}{ffffff}
\definecolor{c5}{HTML}{ffffff}
\definecolor{c6}{HTML}{fffdfd}
\definecolor{c7}{HTML}{f5cfcf}
\definecolor{c8}{HTML}{fffbfb}
\definecolor{c9}{HTML}{ffffff}
\definecolor{c10}{HTML}{fffdfd}
\definecolor{c11}{HTML}{fefafa}
\definecolor{c12}{HTML}{fef7f7}
\definecolor{c13}{HTML}{ffffff}
\definecolor{c14}{HTML}{fffefe}
\definecolor{c15}{HTML}{ffffff}
\definecolor{c16}{HTML}{fefafa}
\definecolor{c17}{HTML}{fdf3f3}
\definecolor{c18}{HTML}{fffefe}
\definecolor{c19}{HTML}{fdf5f5}
\definecolor{c20}{HTML}{ffffff}

\makeatletter
\newcommand{\linebreakand}{%
  \end{@IEEEauthorhalign}
  \hfill\mbox{}\par
  \mbox{}\hfill\begin{@IEEEauthorhalign}
}
\makeatother

\begin{document}

\title{Step-Aligned Script Generation with Verification and Recovery for Web Automation Testing}



\author{
\begin{tabular}{@{}c@{\hspace{1cm}}c@{}}
    \begin{tabular}[t]{@{}c@{}}
        Hridya Dhulipala \\
        \textit{Department of Computer Science} \\
        \textit{The University of Texas at Dallas} \\
        Dallas, USA \\
        hridya.dhulipala@utdallas.edu
    \end{tabular}
    &
    \begin{tabular}[t]{@{}c@{}}
        Xiaokai Rong \\
        \textit{Department of Computer Science} \\
        \textit{The University of Texas at Dallas} \\
        Dallas, USA \\
        xiaokai.rong@utdallas.edu
    \end{tabular}
    \\[6em]
    \begin{tabular}[t]{@{}c@{}}
        Aashish Yadavally \\
        \textit{Department of Computer Science} \\
        \textit{The University of Central Florida} \\
        Orlando, USA \\
        aashish.yadavally@ucf.edu
    \end{tabular}
    &
    \begin{tabular}[t]{@{}c@{}}
        Tien N. Nguyen \\
        \textit{Department of Computer Science} \\
        \textit{The University of Texas at Dallas} \\
        Dallas, USA \\
        tien.n.nguyen@utdallas.edu
    \end{tabular}
\end{tabular}
}


\maketitle

\begin{abstract}
\input{sections/abstract}
\end{abstract}

\input{sections/introduction-ase}
\input{sections/motiv}

\input{sections/formulation}

\input{sections/algo}
\input{sections/emperical_eval_ase}

\input{sections/rq2-new}

\input{sections/rq1-results-new}



\input{sections/rq5}

\input{sections/rq6}
\input{sections/rq3}


\input{sections/threats}

\input{sections/related}
\input{sections/conclusion}

{\bf Data Availability}. Data/code are publicly available~\cite{tessara-website}.





\bibliographystyle{IEEEtran}

\bibliography{sections/references-predex, sections/references-visidroid, sections/references, references-hxagent}

\end{document}

%% file: sections/abstract.tex
{\color{custom-blue2}Automated generation of end-to-end web tests remains costly: developers must identify stable selectors, encode multi-step workflows, attach assertions, and maintain scripts as interfaces evolve. Recent LLM-based web agents can complete natural-language tasks in a live browser session, but task completion is not the same as producing a testware artifact - code that is executable, replayable from a reset state, inspectable by developers, and tied to an oracle. We present {\tool}, an adaptive test-script generation framework that converts natural-language test intent into Playwright scripts consisting of executable interaction code, goal-state assertions, and step-level provenance linking each step to its generated code fragment, selector evidence, execution outcome, and repair history. {\tool} grounds each planned step via Step-Aligned Scripting (SAS), enabling fine-grained failure localization, selective regeneration of failing suffixes, and reversibility-aware backtracking that avoids undoing likely commit points (e.g., form submissions). A three-level verification pipeline (syntactic, execution, and semantic/oracle), detects and repairs malformed code, interaction failures, and incorrect landing states. We evaluate {\tool} on three public web benchmarks, measuring script executability, oracle satisfaction, replay stability, and repair efficiency, and show substantial improvements over baselines and underlying VLMs used directly as script generators.}

%% file: sections/introduction-ase.tex
\section{Introduction}
\label{sec:introduction}

{\color{custom-blue2}

Web applications are central to modern software systems, and end-to-end (E2E) testing is a standard part of their quality-assurance pipeline. E2E tests encode user-visible requirements as executable browser interactions (logging in, searching, filling forms, navigating workflows) and are routinely run in CI/CD pipelines to catch regressions before release. Writing such tests, however, remains expensive and brittle: developers must manually identify stable UI elements, encode the interaction sequence, attach assertions that check the test actually validates the intended behavior, and revise scripts whenever dynamic interface behavior breaks them failures~\cite{9576476}. These costs are amplified in modern web applications, where dynamic DOMs, asynchronous loading, and non-standard widgets routinely invalidate scripts written against a snapshot of the UI. Frameworks such as Playwright~\cite{playwright} or Selenium~\cite{katalon} provide the execution substrate, but authoring and maintaining the scripts that run on them is still largely manual.

A second challenge is handling failures in test scripts. In a testing environment, a failure when executing test scripts is diagnostic information about where a script is fragile and what needs to be repaired. Existing approaches to web interaction largely treat failures uniformly ~\cite{nguyen2025guiagentssurvey}. Retrying a step or regenerating an entire trajectory without attributing the failure to a specific action, selector, or assumption about the page state. For a testing tool, this is a missed opportunity. To elaborate, if a step in the script breaks because a selector changed, a developer (or an automated program repair tool) should be able to localize the fix to that step, inspect why it failed, and regenerate only the affected code fragment — preserving the rest of the script as a still-valid regression test. Treating the whole script as atomic discards exactly the information a maintainable test suite needs.

We present {\tool}, an adaptive test-script synthesis framework that converts natural-language testing intent into executable, replayable, and repairable Playwright tests. Given a testing goal, {\tool} first decomposes it into a sequence of high-level interaction steps, then grounds each step into an executable code fragment via Step-Aligned Scripting (SAS), maintaining an explicit, persistent alignment between each semantic step and the code that implements it. This alignment is what makes the output a testing artifact rather than a transcript. It supports step-level failure localization, selective regeneration of only the failing suffix of a script (preserving validated prefix steps as reusable test fragments), and reversibility-aware backtracking that avoids undoing actions with likely durable side effects (e.g., form submissions, payments).

To assess the correctness of a generated test script, {\tool} applies a three-level verification pipeline. At the syntactic level, generated fragments are checked and repaired against compiler/runtime feedback before they are executed on the browser. At the execution level, each fragment is run, monitored and logged, so failures are attributed to the step that caused them rather than the script as a whole. At the semantic level, an oracle check evaluates whether the resulting page state is actually consistent with the testing intent. This addresses a failure specific to testing tools, where a script can execute without crashing yet land on the wrong page, producing a script that always "passes" without succeeding.

}

In summary, this paper makes the following contributions:

(1) {\bf{\tool}}, a framework for generating executable web automation scripts from natural-language task descriptions via planning, step-aligned scripting, verification, and recovery.

(2) \textbf{Step-Aligned Scripting (SAS)}, which aligns high-level action steps with executable code fragments, enabling stepwise verification, failure localization, and effective backtracking.

(3) A structured verification and recovery, combining a 3-level stack (syntactic, semantic, and execution) with reversibility-aware backtracking to support targeted repair during script generation.

(4) \textbf{Empirical evaluation}: to show {\tool}'s effectiveness in generating accurate and executable web test scripts.


%% file: sections/motiv.tex
\section{Motivation}
\label{sec:motiv}











\subsection{Motivating Example}

In this section, we present a real-world example to motivate our solution. The testing intent is to use the Google Maps website (https://www.google.com/maps/) to answer the query: {\em "Find the difference in travel time by walking versus cycling from the Empire State Building to Central Park."} {\color{custom-blue2}The goal is to generate a reproducible artifact in the form of a Playwright test script. This scenario exposes two fundamental challenges in automated test script generation.}

{\color{custom-blue2}\underline{First}, generating a stable, replayable test script requires a canonical, selector-grounded interaction sequence (one such sequence shown in Fig.~\ref{fig:motiv-plan}). Existing approaches~\cite{zhou2023webarena,pmlr-v235-zheng24e} are optimized for reaching a correct landing page by any viable path rather than producing a fixed, reproducible script. The action sequences these approaches produce compound the problem: natural-language descriptions (e.g., "Click the Search button") cannot be executed directly by Playwright or Selenium without an error-prone conversion step~\cite{pmlr-v235-zheng24e}. Pixel-coordinate representations~\cite{droidagent,autodroid,cheng-etal-2024-seeclick,hong2024cogagentvisuallanguagemodel} execute in a single session but break silently when the UI is updated or rendered across environments, with no indication of which step failed or what needs repair.}


{\color{custom-blue2}\underline{Second}, test scripts must not only be executable, but also satisfy the oracle. They must verify that the expected behavior occurred, rather than merely confirming that execution completed without crashing. Existing agents lack an explicit mechanism to verify that the final GUI state matches the expected behavior, and instead treat successful execution as a proxy for test correctness.}

{\color{custom-blue2}\begin{Observation}
Automated test script generation requires selector-grounded, step-traceable, replayable interaction sequences. Action representations that are ambiguous or coordinate-dependent are insufficient for stable regression testing because they break silently under UI changes and provide no basis for localized repair.
\end{Observation}}


{\color{custom-blue2}\underline{Motivation.} To motivate our solution, we evaluated SeeAct~\cite{pmlr-v235-zheng24e} on the Google Maps testing task above. SeeAct uses GPT-4V with ReAct-style reasoning~\cite{yao2023react} to generate browser actions and executes them through Playwright, interleaving visual observation, reasoning, and UI interaction.}

\begin{figure}[t]
\begin{minipage}{0.4\textwidth}
\centering
  \includegraphics[width=3.4in]{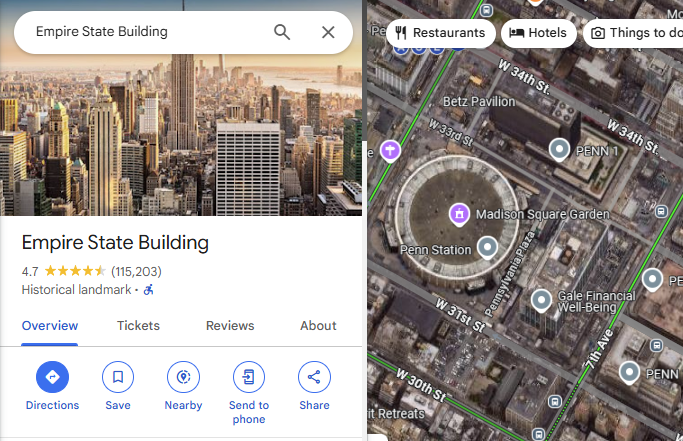}
  \vspace{-6pt}
  \caption{An example of a task on Google Maps}
  \vspace{4pt}
  \label{fig:motiv-1}
\end{minipage}
\hfill








\begin{minipage}[t]{0.48\textwidth}
\centering

\fbox{%
\begin{minipage}{0.95\linewidth}
\scriptsize
\setlength{\parskip}{2pt}

\textbf{Task Plan}

Step 1. Click search bar $\rightarrow$ Type ``Empire State Building, NY''

Step 2. Press Enter $\rightarrow$ Submit search

Step 3. Click ``Directions'' $\rightarrow$ Open directions

Step 4. Click destination bar $\rightarrow$ Type ``Central Park, NY''

Step 5. Click Walking icon $\rightarrow$ Select walking route

Step 6. Record walking time

Step 7. Click Cycling icon $\rightarrow$ Select cycling route

Step 8. Record cycling time

Step 9. Compute time difference
\end{minipage}
}

\vspace{2pt}
\caption{Plan generated by GPT-4V with ReAct.}
\label{fig:motiv-plan}
\end{minipage}
\end{figure}




On the first attempt, GPT-4V completed Steps 1 and 2 but fails at Step 3, as it doesn't locate the ‘Directions’ button using only the DOM tree. After adding the screenshot (Fig.~\ref{fig:motiv-1}), it successfully completes Step 3. However, after entering ‘Central Park, New York,’ SeeAct misclicks the ‘Walking’ icon instead of the search icon, triggering alternative routes and making Step 6 (‘Note walking time…’) ambiguous. It then repeatedly fails to extract the walking time and eventually selects an incorrect button. Thus, SeeAct produces an artifact that would fail the oracle on every subsequent replay, with no way to identify which step introduced the error or repair only that suffix.


\begin{Observation}
    When an incorrect interaction corrupts the UI state mid-script, existing approaches have no mechanism for step-level failure localization or selective suffix repair. The entire script must be regenerated, discarding validated prefix steps and producing no diagnostic information about where the test is fragile.
\end{Observation}

\subsection{Key Ideas}
\label{sec:key-ideas}



\textbf{Key Idea 1. [Step-Aligned Scripting for failure localization and adaptive script generation]}
{\tool} supports failure localization and adaptive script generation through Chain-of-Thought (CoT) planning and Step-Aligned Scripting ({\model}) with verification and backtracking.

Step-level failure localization involves the ability to localize failures to specific reasoning steps. {\tool} achieves this through stepwise verification, where each step in the generated plan is individually grounded and evaluated during execution.

To enable this, {\tool} first uses CoT prompting to generate a high-level sequence of interaction steps. It then applies {\model} to translate the plan into a complete Playwright script while preserving an explicit alignment between each reasoning step and its corresponding code fragment. This alignment allows execution outcomes to be traced back to specific steps. During execution, Playwright interacts with the target application (e.g., Google Maps), and failures are localized to the corresponding step (e.g., Step 6 in Fig. 3).

Stepwise verification further enables adaptive recovery. When a step repeatedly fails, {\tool} backtracks to the most recent successful step and replans from that point using the task description, updated page state, and current URL. It regenerates only the remaining suffix of the plan and incrementally updates the script, preserving prior successful steps. Backtracking is invoked only for actions deemed reversible; otherwise, it falls back to full replanning.

\begin{figure}[htbp]
\centering

\begin{minipage}[t]{0.47\textwidth}
\centering
\scriptsize

\lstset{
    mathescape=true,
    backgroundcolor=\color{lightgray},
    basicstyle=\scriptsize\ttfamily,
    numbers=left,
    numberstyle=\scriptsize\color{black!60},
    keywordstyle=\color{blue!70}\bfseries,
    commentstyle=\color{gray},
    xleftmargin=1em,
    framexleftmargin=1em,
    numbersep=4pt,
    aboveskip=0.3em,
    belowskip=0.3em,
    breaklines=true,
}

\begin{lstlisting}
# Step 1: Type "Empire State Building, NY"
search_box = page.locator("#searchboxinput")
search_box.fill("Empire State Building, NY")

# Step 2: Press Enter
search_box.press("Enter")

# Step 3: Click "Directions"
page.locator("button[data-value*='Directions']").click()

# Step 4: ...
\end{lstlisting}

\end{minipage}\hfill

\vspace{2pt}
\caption{Plan generated by GPT-4V with ReAct.}
\vspace{2pt}
\label{fig:motiv-plan}











\begin{minipage}[t]{0.50\textwidth}

\fbox{%
\begin{minipage}{0.93\linewidth}
\scriptsize
\setlength{\parskip}{2pt}

\textbf{Plan}

1. Search bar $\rightarrow$ Type ``Empire State Building, New York''

2. Press \texttt{Enter} $\rightarrow$ Submit search

3. Click ``Directions''

4. Destination bar $\rightarrow$ Type ``Central Park, New York''

5. Click Search

6. Record walking time

7. Record cycling time

8. Compute the time difference
\end{minipage}
}

\label{fig:motiv-plan}
\end{minipage}

\caption{Step-Aligned Scripting (SAS): mapping between high-level interaction steps and executable Playwright code.}
\label{fig:sas}
\end{figure}


\textbf{Key Idea 2. [3-tier verification and recovery to reduce full replanning]}{\tool} combines syntactic, execution-level, and semantic verification with reversibility-aware backtracking. This allows it to detect (and recover from) malformed code, failed interactions, and incorrect landing states.


\textbf{Key Idea 3. [Automated Test Script Generation]} {\tool} represents an action on a Web page as a 3-tuple of (l,e,a) (l: label, e: Web GUI element, a: operation) that can be automatically transformed into test scripts to support automated test script generation.

%% file: sections/formulation.tex
\section{Formulation}
\label{sec:formulation}

The core problem addressed in this work is as follows. \textbf{Given:} {\color{custom-blue2}a natural-language testing intent and access to a web application under test.}


\textbf{Produce:} {\color{custom-blue2}a replayable test artifact for web automation whose oracle confirms the intended behavior.}

Formally, let \( g \) denote the task goal described in natural language. Let \( W_0 \) denote the initial state of the Web application, comprising both the rendered webpage (visual screenshot) and its Document Object Model (DOM). 
{\color{custom-blue2}The objective is to synthesize an executable interaction script \( C \), with its aligned action sequence \( A = [a_1, a_2, ..., a_n] \) such that executing \( C \) from the initial state \(W_0\) reaches a state \(W_n\) consistent with the goal \(g\). Each action \(a_i \in A\) is represented as a tuple:}
$$
a_i = \langle \code{label}_i, \code{element}_i, \code{action}_i \rangle
$$
\begin{itemize}
  \item \(\code{label}_i\): a semantic description of the step (e.g., ``Click Directions button''),
  \item \(\code{element}_i\): the type of GUI element to be interacted with (e.g., button, textbox, icon),
  \item \(\code{action}_i\): the concrete interaction to perform (e.g., click, hover, type text, extract value).
\end{itemize}




{\color{custom-blue2}{\tool} formulates testware synthesis as a two-stage reasoning-and-execution process, leveraging both natural language reasoning (via CoT prompting) and programmatic execution (via prompting).}

We define the formal behavior as a sequence of state transitions, modeled as:
$
\langle g, s, u, R, C, i \rangle:
$
\begin{itemize}
  \item $g$ is the natural-language goal description,
  \item $s$ is the current screenshot of the Web page,
  \item $d$ is the live filtered DOM of the page,
  \item $u$ is the current URL,
  \item $R = [r_1, \ldots, r_n]$ is the list of reasoning steps (CoT),
  \item $C = [c_1, \ldots, c_n]$ is the list of code fragments ({\model}),
  \item $i$ is the current execution index.
\end{itemize}
The filtered DOM retains only elements whose tag, role, text, or accessibility attributes are compatible with the current reasoning step, e.g., by matching visible text, `aria-label`, `name`, `placeholder`, `title`, and HTML/ARIA element type.


\subsection{Planning Stage (Semantic Abstraction with Planner)}

Chain-of-Thought (CoT) prompting is used to generate an explicit sequence of reasoning steps \( R = [r_1, r_2, ..., r_n] \) that outline how to complete the task \( g \). Each step \( r_i \) describes a conceptual action on the GUI (e.g., {\em ``Type destination into the search bar''}). Inputs include \( g \), screenshot of \( s_0 \), and a filtered DOM $d$.
Given a goal $g$, a screenshot $s$, and a URL $u$, the Planner produces reasoning steps:
\[
{\langle g, s, d, u,  \varnothing, \varnothing, 0 \rangle \xrightarrow{\code{plan}} \langle g, s, d, u, R, \varnothing, 0 \rangle}
\]

\subsection{Scripting Stage (Programmatic Grounding with Scripter)}

Our Step-aligned Scripting ({\model}) prompting translates each reasoning step \( r_i \) into an executable code fragment \( c_i \) to perform the desired action. This results in a complete script \( C = [c_1, c_2, ..., c_n] \), where each code fragment is aligned to a reasoning step.
Given reasoning steps $R = [r_1, \ldots, r_n]$, the Scripter (using {\model}) produces corresponding code fragments:
\[
{ \langle g, s, d, u, R, \varnothing, 0 \rangle \xrightarrow{\code{script}} \langle g, s, d, u, R, C, 0 \rangle }
\]
\subsection{Execution \& Stepwise Verification}
\label{sec:verification}

{\tool} executes the script using Playwright to interact with the website. Observing execution success/failure at each \( c_i \) enables stepwise verification of \( r_i \). Let \( V(C) \) denote the verification function that executes the script~\( C\) = \([c_1, c_2, ..., c_n] \) and monitors the outcome. The function \( V \) returns a vector of execution results \( [v_1, v_2, ..., v_n] \), where each \( v_i\) \(\in \{ \code{success}, \code{failure} \} = \{\checkmark, \times\}\) reflects the execution~result of the code fragment \( c_i \). The value of \( v_i \) serves as a proxy to verify the correctness of the respective reasoning step \( r_i \) in the plan.
This alignment enables {\em verification}: if \( v_i\) = \( \times \), the reasoning step \( r_i \) is considered incorrect or incomplete. Our verification has three levels:

(1) {\bf \underline{Syntactic} level}: if the generated Playwright code contains syntax errors, {\tool} provides the corresponding error message produced by the compiler as feedback to the Scripter LLM to repeatedly re-prompt for correction until the output $C$ is executable or time-out (complete~replan).


(2) {\bf \underline{Execution} level}: The execution verification function \( V(C) \) executes code $C$
\& returns a vector of outcomes \( v_i \in \{\checkmark, \times\} \), where each \( v_i \) is the result of executing \( c_i \). Then, stepwise verification aligns each \( r_i \) with \( c_i \) via:
$
V(C) = [v_1, \ldots, v_n] \quad \text{where } v_i = \code{exec}(c_i)
$
\[
\langle g,s,d,u,R,C,0\rangle
\xrightarrow{\code{ver}}
\begin{cases}
\text{Success},
&
\begin{aligned}[t]
&\forall i,
&v_i=\checkmark
\end{aligned}
\\[4pt]
\text{Fail at } i^*,
&
\begin{aligned}[t]
&v_{i^*}=\times,\\
&\forall j<i^*,\; v_j=\checkmark
\end{aligned}
\end{cases}
\]

This provides the foundation for \textbf{step-level failure localization}, allowing the agent to identify and isolate the specific reasoning step that caused the failure. 




(3) {\bf  \underline{Semantic} level (oracle evaluation)}: {\color{custom-blue2}The LLM-based check is treated as an oracle. It assesses whether the resulting page/state satisfies the task intent g, rather than merely confirming that execution did not crash. After executing the prefix ([$c_1,\ldots,c_n$]), we capture the last observed page/state ($\mathcal{W}_n$) nd feed into an LLM the task description (g) and the landing page. The LLM outputs a binary judgment ($\hat{v}^{sem}_n\in{\checkmark,\times}$) indicating whether the oracle is satisfied or not with the final state of the script execution.}


\subsection{Adaptation via Partial Replanning}
If execution fails at step \( c_k \), \tool{} marks the respective reasoning step \( r_k \) as failed and initiates partial replanning. A new plan \( R' = [r_k', r_{k+1}', ..., r_m'] \) is generated using the goal \( g \), the latest state \( s \), and an updated screenshot and filtered DOM \(d\). The new script fragments \( [c_k', ..., c_m'] \) are appended to the preserved prefix \( [c_1, ..., c_{k-1}] \) of the previous script. We use ``{\bf adaptation}'' for such partial~replanning:

$$
\frac{v_{k} = \times}{\langle g, s, d, u, R, C, k \rangle \xrightarrow{\code{replan}} \langle g, s', d', u', R', C', k \rangle}
$$ $s' = \code{GetScreenshot()}$, $u' = \code{GetCurrentURL()}$, $R' = [r'_{k}, \ldots,   r'_m]$, $d' = \code{GetUpdatedfilteredDOM()}$, (new CoT steps), $C' = [c'_{k}, \ldots, c'_m]$ (new {\model} fragments).

\subsection{Backtracking Strategy and Procedure}

\subsubsection{GUI elements in Web applications.}
We define a Web GUI element as an \emph{interactable} object exposed by the page's DOM (Document Object Model) and accessibility semantics. Concretely, we type an element $e$ by (i) its \textbf{HTML category} (e.g., form controls such as \code{input}, \code{select}, \code{textarea}, \code{button}, and navigation targets such as \code{a}), and (ii) its \textbf{WAI-ARIA role} (e.g., \code{button}, \code{link}, \code{textbox}, \code{checkbox}, \code{radio}, \code{combobox}, \code{tab}, \code{menuitem}, \code{dialog}). These standards provide GUI element types that can be obtained from DOM attributes and/or the accessibility tree exposed by user agents to assistive technologies~\cite{whatwg_html_forms,whatwg_html_button,wai_aria_12,wai_aria_usage,wai_aria_uai}.

\subsubsection{Reversible vs irreversible actions.} Let the application state be $S$ (including URL, rendered UI/DOM, and relevant server/session state). An action $a$ applied to element $e$ induces a transition $S \xrightarrow{a(e)} S'$. We call $a(e)$ \textbf{reversible} if there exists a \emph{bounded} sequence of backtrack actions $\pi$ (e.g., close dialog, navigate back, reset filters, clear inputs) such that $S' \xrightarrow{\pi} \tilde{S}$ and $\tilde{S} \equiv S$ under an observational equivalence (same page context/mode and no persistent side effects). Conversely, $a(e)$ is \textbf{irreversible} if no such bounded $\pi$ exists without external intervention, privileged operations, or non-guaranteed compensation. This distinction aligns with the Web workflow notion of a \emph{commit point}: actions that submit data to the server (e.g., form submission initiated by a submit button) typically create durable effects and are treated as irreversible by default~\cite{whatwg_html_forms,whatwg_html_button}.

\subsubsection{Categorization by element types (typical cases).}
Using the HTML/ARIA element taxonomy above, we categorize common Web actions as follows.

\noindent\textbf{\underline{Reversible actions} (safe to backtrack)} are defined concretely based on the following scenarios:

\textbf{\em View/navigation without commit:} clicking \code{link}/\code{tab}/\code{menuitem}, opening/closing panels, switching tabs/sections, and dismissing \code{dialog} overlays (e.g., \code{Esc} or close button)---these primarily change the view state and are typically undoable by ``Back/Close/Return-to-home'' operations~\cite{wai_aria_12,wai_aria_usage}.

\textbf{\em Pre-commit form editing:} typing into \code{textbox} (e.g., \code{input}, \code{textarea}), changing selections in \code{combobox}/\code{listbox} (e.g., \code{select}), and toggling \code{checkbox}/\code{radio} \emph{before} an explicit save/submit---these can be reversed via clearing inputs, resetting the form, or reverting UI selections~\cite{whatwg_html_forms,wai_aria_12}.

\textbf{Pure UI interactions:} scrolling, expanding/collapsing menus, focus/hover effects that don't commit server-side state.

\noindent\textbf{\underline{Irreversible actions} (avoid backtracking; re-plan from last safe checkpoint):}

\textbf{\em Commit actions:} clicking a confirmation/\code{button} that triggers \emph{form submission} or otherwise persists changes (e.g., ``Save'', ``Submit'', ``Send''). In HTML, submit buttons are associated with initiating form submission, making them natural commit points for durable state changes~\cite{whatwg_html_button,whatwg_html_forms}.

\textbf{\em Destructive or externally impactful actions:} permanent deletion, sending messages or notifications, initiating payments or transfers, and admin/global operations (e.g., applying policies), as these produce durable side effects that cannot be reliably undone through bounded backtracking.



{\bf Formulation.} If the partial adaptive replanning fails to run after \( M \) attempts, {\tool} engages a \textit{backtracking procedure}. Let \( j < k \) be~the index of the last known successful step. It performs:
  \begin{itemize}
    \item Reverts the Web app to the visual/structural state \( s_j \) via operations in {\bf reversible actions}. If the operation is {\em irreversible, backtracking is stopped and complete replanning is initiated},
    \item Rolls back the execution trace to step \( c_j \),
    \item Triggers a fresh replanning from step \( r_{j+1} \) onward:
    $
    R'' = [r_{j+1}'', r_{j+2}'', ..., r_n''],
    C'' = [c_{j+1}'', c_{j+2}'', ..., c_n'']
    $
  \end{itemize}

This process repeats recursively until either the task succeeds or the maximum attempts to restore an observationally equivalent safe UI state is reached. This mechanism enables \tool{} to explore alternate paths that may avoid GUI anomalies, misidentifications, or transient layout issues that caused the original failure.

If re-planning fails repeatedly at \( k \), then backtrack to prior index (\( k - 1 \)):
\[
\frac{\text{retries}(k) > \delta}{\langle g, s, d, u, R, C, k \rangle \xrightarrow{\code{backtrack}} \langle g, s, d, u, R, C, k - 1 \rangle}
\]

and re-enter the replanning rule from (\( k - 1 \)). 

\begin{figure*}[t]
  \centering
  \includegraphics[width=5.6in]{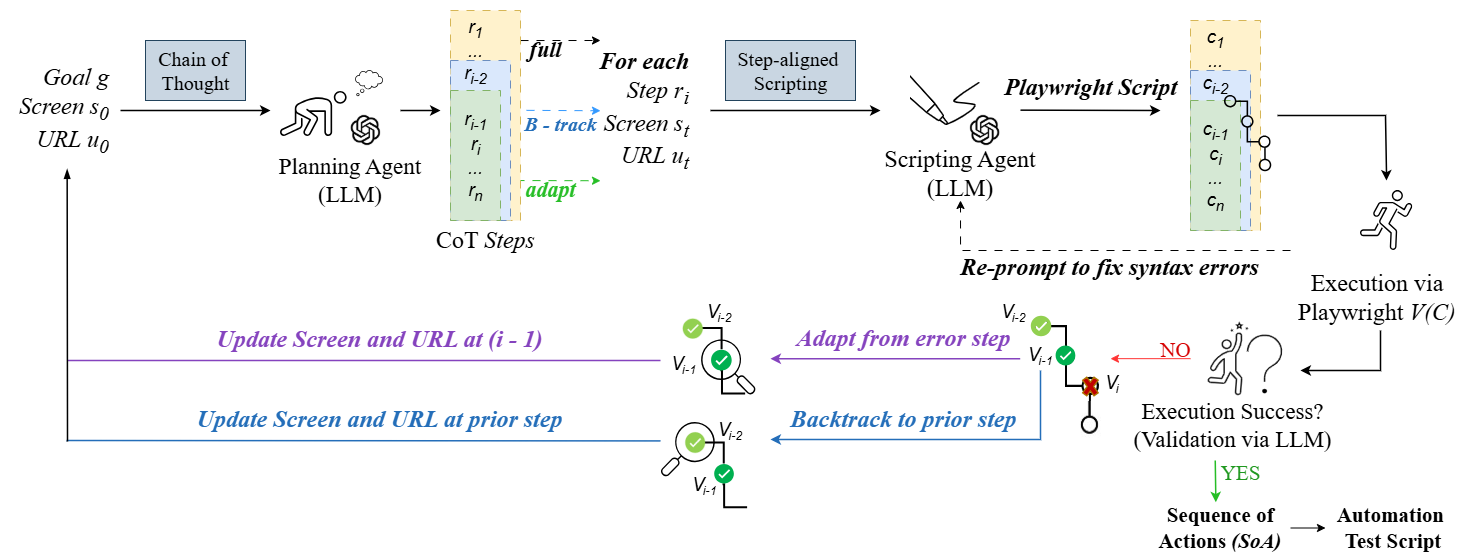} 
  \vspace{-6pt}
  \caption{{\tool}: Adaptive Planning for Test Script Generation for Web Automation}
  \vspace{-8pt}
  \label{fig:arch}
\end{figure*}

{\bf Termination}. The process terminates when all code fragments succeed or a failure threshold is reached: 

(1) $\forall i,\, v_i = \checkmark \quad \Rightarrow \quad \code{return } C$, and 

(2) $\code{failures exceeded} \quad \Rightarrow \quad \code{return } \bot$.


%% file: sections/algo.tex
\section{{\tool} Algorithm}
\label{sec:algo}

This section presents the algorithm behind {\tool}'s workflow illustrated in Fig.~\ref{fig:arch} (Algorithm~\ref{alg:atlas}).

\underline{Initialization}: The {\tool} agent begins with three inputs: a natural language task goal $g$, an initial webpage screenshot $s_0$ (captured via Playwright), and the initial page URL $u_0$. These inputs represent the agent’s initial perceptual and contextual state. {\tool} initializes its internal variables, setting the timestep counter $t$ to zero and initializing the reasoning trace $R$ and the execution script $C$ as empty lists. These structures will be populated during subsequent stages of the algorithm.

\underline{\bf Step 1}: {\bf Planning via Chain-of-Thought}: In the planning stage, the agent decomposes the overall task $g$ into a sequence of high-level reasoning steps $R = [r_1, r_2, \dots, r_n]$. This is achieved through CoT prompting, conditioned on the current goal and the screenshot $s_t$ at $t$. Each reasoning step $r_i$ captures an abstract subgoal or intended interaction with the web interface (e.g., “select the departure city,” “click on the login button”). This decomposition defines a plan for accomplishing the task and provides the reasoning for subsequent code generation and verification. 

\underline{\bf Step 2}: {\bf Scripting via Step-aligned Scripting ({\model})}: For each reasoning step $r_i$, the agent uses {\model} to generate a code fragment $c_i$ that implements the action via Playwright. The {\model} prompt incorporates $r_i$, the filtered live DOM derived from the current screenshot $s_t$ via BeautifulSoup~\cite{beautifulsoup}, and the current URL $u_t$, helping the model to map text-based intent to contextually grounded executable code. The resulting sequence of code $C = [c_1, \dots, c_n]$ forms a task script aligned stepwise with the reasoning trace $R$. 

\underline{\bf Step 3}: {\bf Execution and Stepwise Verification}: The entire script $C$ is executed sequentially in the browser environment using Playwright. Specifically, {\tool} applies a verification function $V$ that evaluates the execution success/failure of code fragment $c_i$. This yields a result vector $[v_1, \dots, v_n]$, where $v_i \in \{\code{success}, \code{failure}\}$ reflects the outcome of executing step $c_i$. By~attributing verification result $v_i$ back to its associated reasoning step $r_i$, the agent achieves {\em failure localization/verification}.

\input{sections/pseudo-code}

\underline{\bf Step 4}: {\bf Adaptation and Backtracking}: If any verification result $v_i$ indicates failure, {\tool} enters an adaptation loop designed to recover from errors. The agent identifies the first failed step $i^*$ and checks whether it has exhausted its retry budget for $r_{i^*}$. If so, it backtracks to the preceding step $r_{i^*-1}$, indicating that earlier reasoning may be reconsidered. It then captures the current state, including the updated screenshot $s_{t+1}$ and URL $u_{t+1}$, and re-invokes the CoT planner to regenerate a new set of reasoning steps $R_{\text{new}}$ beginning from the backtracked point. If the operation is irreversible, the backtrack process is halted. Else, for each new step, code fragments are generated via {\model} prompting, and the failing suffix of the original plan and script is repaired. The updated script $C$ is re-executed and re-verified. This loop continues until the script succeeds or retry/backtracking limits are reached. 

Termination: {\tool} terminates when all verification steps succeed, yielding a complete executable script $C$. If {\bf all attempts fail} despite backtracking and adaptation, it {\bf replans completely}. This marks the completion of the full test script.

\input{sections/planner-prompt}

%% file: sections/pseudo-code.tex
\begingroup
\setlength{\textfloatsep}{4pt plus 1pt minus 1pt}
\setlength{\floatsep}{4pt plus 1pt minus 1pt}
\setlength{\intextsep}{4pt plus 1pt minus 1pt}
\setlength{\abovecaptionskip}{2pt}
\setlength{\belowcaptionskip}{2pt}

\begin{algorithm}[t]
\scriptsize
\caption{Adaptive Test Script Generation for Web Automation}
\label{alg:atlas}
\textbf{Input:} task goal $g$ in text, initial screenshot $s_0$, initial URL $u_0$ \\
\textbf{Output:} Executable Playwright script $C$ that performs the task
\begin{algorithmic}[1]
\State \textbf{Initialize:} $t \gets 0$, $s_t \gets s_0$, $u_t \gets u_0$, $R \gets []$, $C \gets []$
\Statex \textbf{// Step 1: Planning (via CoT prompting)}
\Statex \textit{// Generate reasoning steps $R = [r_1, r_2, \dots, r_n]$}
\State $R \gets \text{Planner.LLM-CoT}(g, s_t)$
\Statex \textbf{// Step 2: Scripting (via {\model} prompting)}
\ForAll{reasoning step $r_i \in R$}
    \State $c_i \gets \text{Scripter.LLM-{\model}}(r_i, \text{FilteredDOM}(s_t), u_t)$
    \While{$c_i$ has a syntax error}
        \State $c_i \gets \text{Scripter.LLM}(c_i, errormsg)$
    \EndWhile
    \State Append $c_i$ to $C$
\EndFor
\Statex \textbf{// Step 3: Execution \& Stepwise Verification}
\Statex \textit{// Execute script $V(C)$; $v_i \in \{\code{success}, \code{failure}\}$ for $c_i$}
\State $v_1, \dots, v_n \gets V(C)$
\Statex \textbf{// Step 4: Adaptation \& Backtracking}
\While{any $v_i = \texttt{failure}$ and retry limit not exceeded}
    \State $i^* \gets \min\{i : v_i = \code{failure}\}$
    \If{retry attempts at $r_{i^*}$ exhausted}
        \State $i^* \gets i^* - 1$ \Comment{Backtrack if operation $[i^*]$ is reversible}
    \EndIf
    \State $s_{t+1} \gets \text{GetScreenshot}()$; $u_{t+1} \gets \text{GetCurrentURL}()$
    \State $R_{\text{new}} \gets \text{Planner.LLM-CoT}(g, s_{t+1}, \text{resume from } r_{i^*})$
    \ForAll{new reasoning step $r_j \in R_{\text{new}}$}
        \State $c_j \gets \text{Scripter.LLM-{\model}}(r_j, \text{FilteredDOM}(s_{t+1}), u_{t+1})$
    \EndFor
    \State Replace $r_{i^*:}$ in $R$ with $R_{\text{new}}$
    \State Replace $c_{i^*:}$ in $C$ with new $c_j$
    \State $v_1, \dots, v_n \gets V(C)$
    \State $t \gets t + 1$
\EndWhile
\State \Return $C$
\end{algorithmic}
\end{algorithm}
\endgroup

%% file: sections/emperical_eval_ase.tex
\section{Empirical Evaluation}
\label{sec:emperical_eval}

For evaluation, we seek to answer the following~questions:

\noindent {\bf RQ1. [Effectiveness on Test Script Generation].} How effective is {\tool} in generating executable test scripts for the tasks compared to the existing approaches?

{\color{custom-blue2} \noindent {\bf RQ2. [Plan Fidelity].} How faithfully does {\tool} generate the underlying interaction sequence that the test script must encode, compared to existing approaches?}


\noindent {\bf RQ3. [Analysis on Backtracking].} How well does our backtracking improve the performance?

\noindent {\bf RQ4. [Accuracy of Test Oracle Generation].} How accurate is our three-level stepwise verification?

\noindent {\bf RQ5. [Ablation Study].} How do different components in {\tool} contribute to its performance?

\textbf{Datasets.} {\em \textbf{WebVoyager}}~\cite{he2024webvoyager} has 643 task queries on visual and HTML texts in 15 websites. {\em \textbf{VisualWebArena}} \cite{visualwebarena2024} has 910 visually grounded tasks, including a subset of the original WebArena. {\em \textbf{WorkArena}} \cite{drouin2024workarena} targets enterprise tasks in a~ServiceNow instance. We chose a representative set of 341 tasks. 

\textbf{Baselines.} {\color{custom-blue2}We evaluate test script generation using GPT-4V, GPT-4oV, Gemini-Pro, Llama 3, and Mixtral as the underlying reasoning models. We selected SeeAct~\cite{pmlr-v235-zheng24e} as the primary baseline because it is the closest prior work, translating natural-language intents into executable sequences of UI interactions. Other web agents lack an equivalent mechanism for producing executable action sequences, making direct comparison less meaningful and introducing implementation-dependent biases. To ensure a fair comparison, both {\tool} and SeeAct use GPT-4V as the reasoning model. Section~\ref{sec:related} explains our rationales for baseline choices.}

\textbf{Procedure.} Each data sample consists of a start URL and a textual task description. To reduce non-determinism, for each sample, the models were allowed to run for a maximum of 10 minutes, with additional termination criteria: a maximum of 20 invalid actions or executions for SeeAct and {\tool}. At the end of each run, both models produced their respective action sequences; {\tool} additionally generated the Playwright script derived from SoAs. For each task, an annotator executed and validated a successful path from the benchmark’s initial state while a logger recorded each action. If multiple valid paths existed, we retained a shortest validated successful path as the canonical SoA. {\color{custom-blue2}Each experiment was run 3 times, and the mean performance was reported. }

\textbf{Metrics}. For evaluation, we use the following four metrics:


{\em \textbf{(1) Script Executability (SE)}}:  {\color{custom-blue2}A test script is considered executable if the automation run reaches the assumed final state without crashing or exceeding the time budget.}


{\color{custom-blue2}{\em \textbf{(2) Oracle Satisfaction (OS)}}: This is the percentage of tasks where the test script both executes successfully and satisfies its oracle — i.e., the landing state is judged consistent with the testing intent. Oracle Satisfaction is the metric we treat as evidence that a generated script functions as a meaningful test rather than merely an executable one}

{\em \textbf{(3) Prefix-match (PM)}}. Prefix-match quantifies how much of a synthesized script survives as a valid, reusable test fragment even when the full sequence is not exactly correct. It is calculated as the proportion of correct actions from the beginning of the sequence up to the first incorrect action. This corresponds to the proportion of the grounded script's prefix that requires no developer correction.


{\em \textbf{(4) Exact-match (EM)}}.A generated web automation script is an exact match if every code fragment correctly interacts with the intended Web element in the correct order and the script successfully completes the specified task. Such scripts require no human correction, representing the upper bound of zero-touch test generation.
Exact-match tasks are a subset of prefix-match ones, which are a subset of successful ones, which are a subset of completed~ones.




{\textbf{Complexity}}. {\em Interface complexity} is decided by {\bf(i)} the number of pages a task requires, {\bf(ii)} the density of actionable UIs (e.g., buttons, dropdowns), and {\bf(iii)} the clarity of interface elements (text-labeled or icon-based). 
%
{\em Task complexity} is driven by {\bf(i)} the length of the optimal reasoning or navigation sequence, and {\bf(ii)} computational reasoning, e.g., aggregation, filtering, or~comparison. 

%% file: sections/rq2-new.tex
\subsection{Effectiveness on Test Script Generation (RQ1)}
\label{sec:rq2-new}

\input{sections/rq2-table}

{\color{custom-blue2}We evaluate whether {\tool}'s generated scripts are executable and oracle-satisfying as standalone testware artifacts comparing {\tool} with GPT-4V as underlying LLM against multimodal VLMs prompted to complete the tasks using Chain-of-Thought~\cite{drouin2024workarena,wei2023chainofthoughtpromptingelicitsreasoning}.}
For {\tool}, the final script output is executed in Playwright and checked for Script Executability and Oracle Satisfaction. For baselines like SeeAct, we convert its executable action sequences into Playwright scripts and evaluate them under the same execution protocol.

\noindent {\bf Comparison with VLM-based UI Agents (Table~\ref{tab:rq2-vlm})}. The baselines represent the VLMs underlying the CoT UI agents. {\em Since VLMs do not produce executable SoAs, we only compare them in Oracle Satisfaction}. As seen, {\tool} with GPT‑4V performs better than those~VLMs.




{\color{custom-blue2}GPT‑4oV falls to 41.8\% Oracle Satisfaction on {\em WorkArena}, 55.5\% on WebVoyager and 24\% on VisualWebArena; GPT‑3.5 reaches only 6.1\%. Generic VLMs struggle with complex enterprise UIs due to the lack of DOM grounding, state abstraction, and verification. {\tool} overcomes these challenges through stable selector generation and iterative verify–repair. Since test scripts are generated by grounding planned UI steps into executable code, we next evaluate the quality of the underlying action sequences in Section~\ref{sec:rq1}.}



\noindent {\bf Comparison with LLM-based agent SeeAct (Table~\ref{tab:rq2-overall}).}
On WebVoyager, {\tool} achieves a 55.7\% higher Script Executability and 2.15X higher Oracle Satisfaction than SeeAct. On VisualWebArena, the Script Execution gain is by 65.1\% and that of Oracle Satisfaction by 2.51x. 

%% file: sections/rq2-table.tex
\begin{table}[t]
\vspace{-8pt}
\centering
\caption{Test Script Generation Effectiveness (RQ1). }

\label{tab:rq2-vlm}
\vspace{-6pt}
\small
\begin{tabular}{|c|l|c|}
\hline
{\bf Dataset} & \multicolumn{1}{c|}{\bf Model} & \multicolumn{1}{c|}{\bf Oracle Satisfaction (\%)} \\ \hline
\multirow{3}{*}{\begin{tabular}[c]{@{}c@{}}Web\\Voyager\end{tabular}} & Claude-3-Opus & 52.8 \\ \cline{2-3}
& GPT-4oV & 55.5 \\ \cline{2-3}
& \textbf{{\tool} (GPT-4V)} & \textbf{59.34} \\ \hline
\multirow{6}{*}{\begin{tabular}[c]{@{}c@{}}Visual\\Web\\Arena\end{tabular}} & Gemini-Pro & 5.71 \\ \cline{2-3}
& Llama3 & 11.0 \\ \cline{2-3}
& Mixtral & 12.6 \\ \cline{2-3}
& GPT-3.5 & 6.7 \\ \cline{2-3}
& GPT-4oV & 24.0 \\ \cline{2-3}
& \textbf{{\tool} (GPT-4V)} & \textbf{54.83} \\ \hline
\multirow{5}{*}{\begin{tabular}[c]{@{}c@{}}Work\\Arena\end{tabular}} & GPT-3.5 & 6.1 \\ \cline{2-3}
& GPT-4oV & 41.8 \\ \cline{2-3}
& Llama3 & 17.9 \\ \cline{2-3}
& Mixtral & 12.4 \\ \cline{2-3}
& \textbf{{\tool} (GPT-4V)} & \textbf{51.9} \\ \hline
\end{tabular}%
\vspace{-10pt}
\end{table}

\newlength{\coltabw}
\setlength{\coltabw}{\dimexpr(\textwidth-\columnsep)/2\relax}

\begin{table*}[t]
\centering
\vspace{-9pt}

\begin{minipage}[t]{\coltabw}
\centering
\caption{Test Script Generation in three datasets (RQ1).}
\label{tab:rq2-overall}
\vspace{-6pt}
\small
\setlength{\tabcolsep}{3pt}
\begin{tabular}{|c|c|c|c|}
\hline
\multirow{2}{*}{\bf Dataset} & \multirow{2}{*}{\bf Model} & \multicolumn{2}{c|}{\bf Accuracy (\%)} \\ \cline{3-4}
                         &                        & \rule{0pt}{2.4ex}Script Executed & Oracle Satisfaction \\ \hline
\multirow{2}{*}{WebVoyager} & SeeAct & {50.23} & 27.49 \\ \cline{2-4}
& \textbf{{\tool}} & {\textbf{78.2}} & \textbf{59.34} \\ \hline
\multirow{2}{*}{VisualWebArena} & SeeAct & {42.71} & 20.02 \\ \cline{2-4}
& \textbf{{\tool}} & {\textbf{66.49}} & \textbf{54.83} \\ \hline
\multirow{2}{*}{WorkArena} & SeeAct & {38.34} & 16.54 \\ \cline{2-4}
& \textbf{{\tool}} & {\textbf{59.77}} & \textbf{51.9} \\ \hline
\end{tabular}
\end{minipage}
\hfill
\begin{minipage}[t]{\coltabw}
\centering
\caption{SoA Generation in three datasets (RQ2).}
\label{tab:rq1-results}
\vspace{-6pt}
\small
\setlength{\tabcolsep}{6pt}
\begin{tabular}{|c|c|cccc|}
\hline
\multirow{2}{*}{\textbf{Dataset}} &
  \multirow{2}{*}{\textbf{Model}} &
  \multicolumn{4}{c|}{\textbf{Accuracy (\%)}} \\ \cline{3-6}
 &
   &
  \multicolumn{1}{c|}{\rule{0pt}{2.4ex}\textbf{SE}} &
  \multicolumn{1}{c|}{\textbf{OS}} &
  \multicolumn{1}{c|}{\textbf{EM}} &
  \textbf{PM} \\ \hline
\multirow{2}{*}{WebVoyager} &
  SeeAct &
  \multicolumn{1}{c|}{51.23} &
  \multicolumn{1}{c|}{47.16} &
  \multicolumn{1}{c|}{43.49} &
  75.02 \\
 &
  \textbf{\tool} &
  \multicolumn{1}{c|}{\textbf{82.33}} &
  \multicolumn{1}{c|}{\textbf{68.56}} &
  \multicolumn{1}{c|}{\textbf{66.56}} &
  \textbf{78.74} \\ \hline
\multirow{2}{*}{VisualWebArena} &
  SeeAct &
  \multicolumn{1}{c|}{44.78} &
  \multicolumn{1}{c|}{41.34} &
  \multicolumn{1}{c|}{33.91} &
  58.86 \\
 &
  \textbf{\tool} &
  \multicolumn{1}{c|}{\textbf{75.17}} &
  \multicolumn{1}{c|}{\textbf{63.26}} &
  \multicolumn{1}{c|}{\textbf{63.52}} &
  \textbf{74.61} \\ \hline
\multirow{2}{*}{WorkArena} &
  SeeAct &
  \multicolumn{1}{c|}{38.34} &
  \multicolumn{1}{c|}{19.93} &
  \multicolumn{1}{c|}{16.54} &
  24.37 \\
 &
  \textbf{\tool} &
  \multicolumn{1}{c|}{\textbf{70.09}} &
  \multicolumn{1}{c|}{\textbf{59.02}} &
  \multicolumn{1}{c|}{\textbf{56.76}} &
  \textbf{62.67} \\ \hline
\end{tabular}
\end{minipage}

\end{table*}

%% file: sections/rq1-results-new.tex
\subsection{Sequences of Action Generation (RQ2)}
\label{sec:rq1}

\begin{figure}
  \centering
  \includegraphics[width=3.5in]{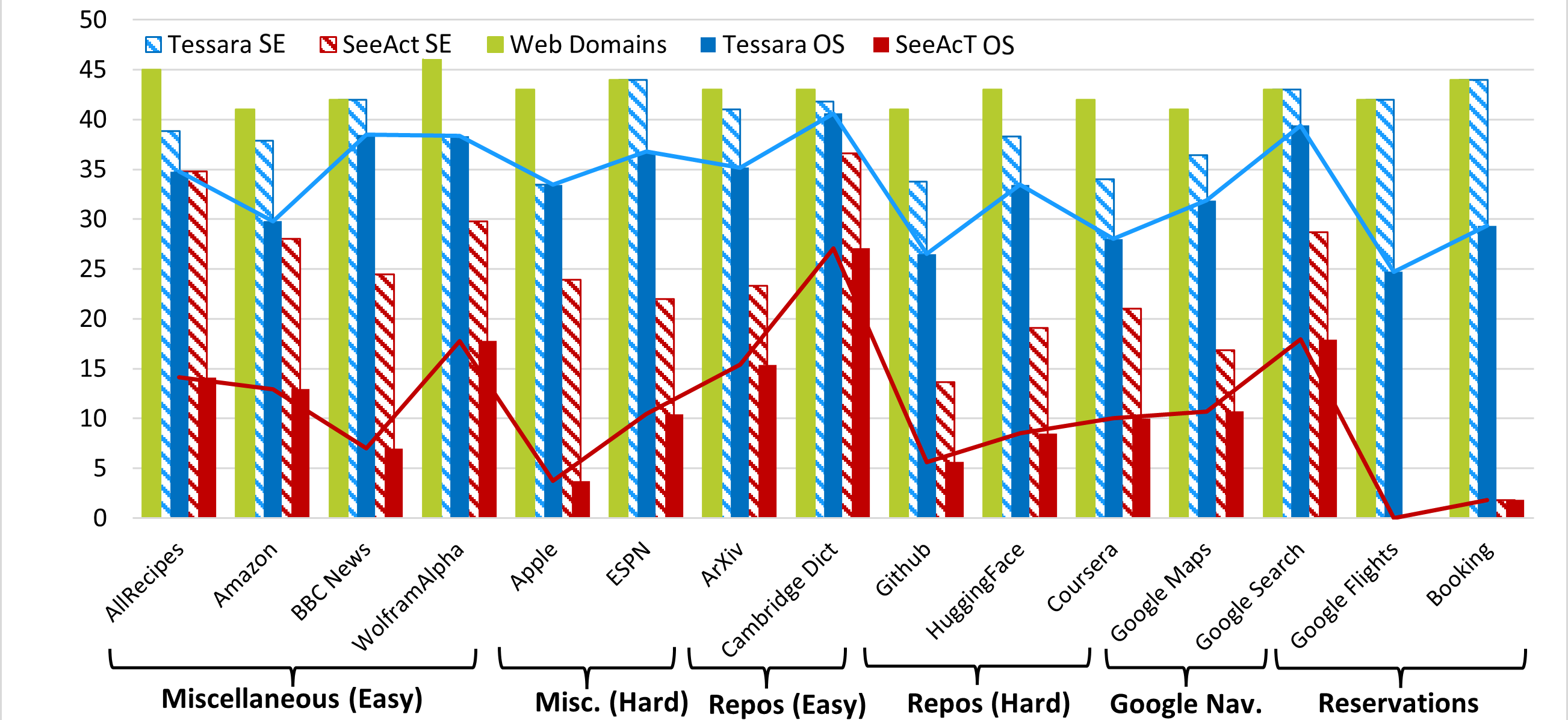}
  \vspace{-4pt}
  \caption{(\underline{SE and OS}) Performance on \underline{WebVoyager} dataset. The bars (green, patterned and solid) show the total unique, completed \& success tasks. Line graphs show trends (RQ2).}
  \vspace{-12pt}
  \label{fig:rq1_webvoyager}
\end{figure}

\textbf{Results on Action Sequence Generation}. On WebVoyager, {\tool} achieves a 60.7\% relative improvement in Script Executability over SeeAct (Table~\ref{tab:rq1-results}). Oracle Satisfaction increases from 47.16\% with SeeAct to 68.56\% with {\tool}, representing 45.4\% improvement of {\tool} over the baseline. On the VisualWebArena benchmark, {\tool} outperforms SeeAct on all four metrics. As shown in Table ~\ref{tab:rq1-results}, {\tool}’s Script Executability improves by 68\% over SeeAct, reflecting greater stability in reaching final states, while its Oracle Satisfaction is 53\% higher. On WorkArena, {\tool} relatively improves over SeeAct by 1.83X on Script Executability and 2.96X on Oracle Satisfaction. 



%

\noindent {\bf Exact-match and Prefix-match Comparison.} {\color{custom-blue2}On WebVoyager, {\tool} achieves Exact-Match for 66.56\% of tasks and Prefix-Match for 78.74\%, meaning the large majority of generated plans require no, or only suffix-level, developer correction before being grounded into a usable test script. SeeAct, by contrast, produces Exact-Match for only 43.49\% of tasks against 75.02\% Prefix-Match — a markedly lower zero-touch rate despite comparable prefix salvageability.}

 
On VisualWebArena, {\tool} outperforms SeeAct with the largest improvement in dynamic/multi-step interfaces (Table~\ref{tab:rq1-vwebarena}), specifically 
%
in composite tasks requiring cross-site navigation. For instance, 
{\em “Finding~the walking distance between Empire State Building and the \#1 public university in NY”} spans Wikipedia and OpenStreetMap, requiring state maintenance, data storage, and validation.
On WorkArena, {\tool} improves over SeeAct by 3.4X on exact-match and 2.57X on prefix match. The largest improvement is in ``List'' tasks (Table~\ref{tab:rq1-workarena}), where form-based retrieval demands precise filtering/sorting and verification of state changes. 
SeeAct’s next-action, vision-first policy is brittle, often misreading UI semantics or falling into~scrolling loops.


\begin{figure}[t]
\begin{minipage}{0.47\textwidth}
   \centering
  \includegraphics[width=2.5in]{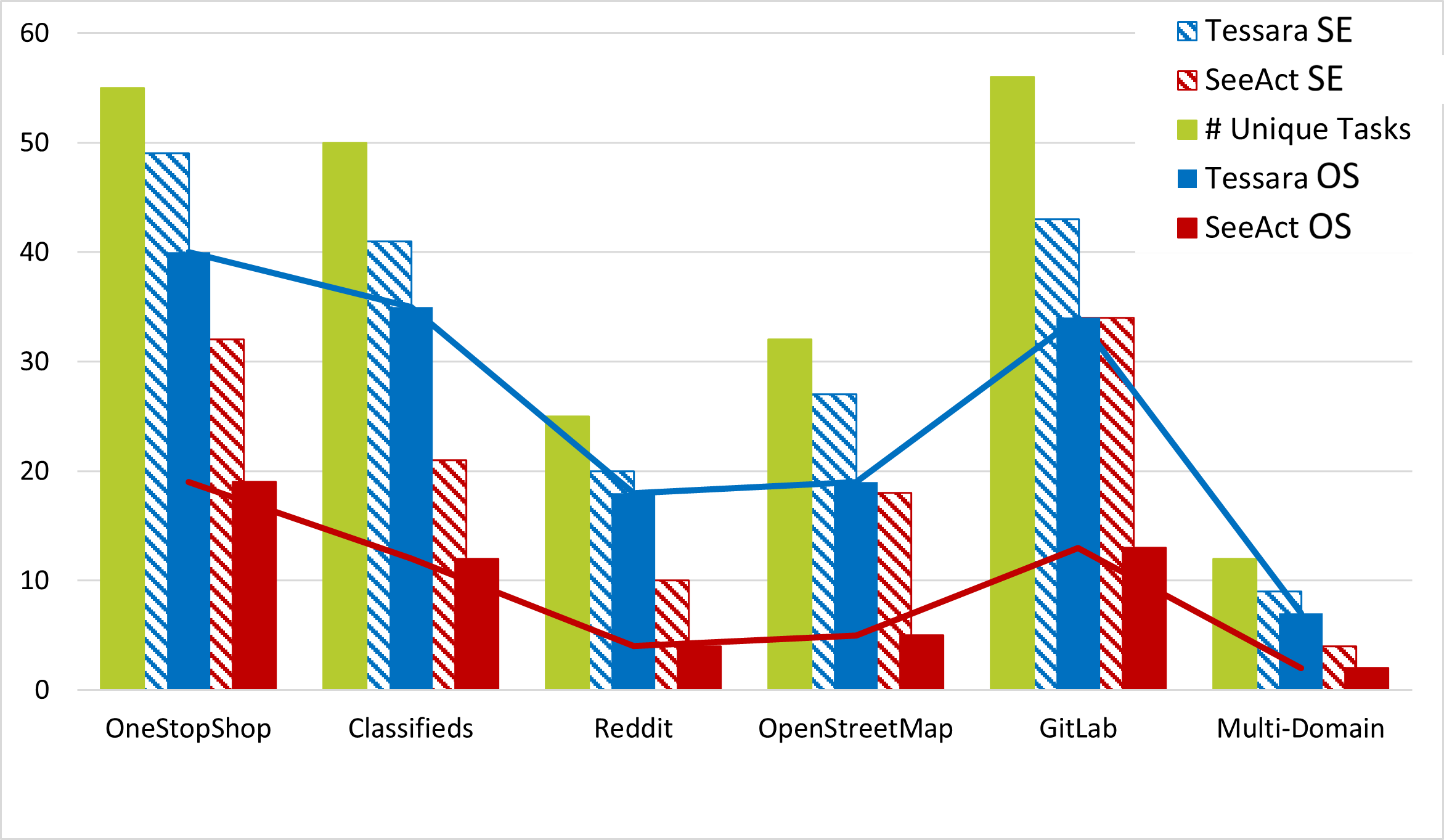}
  \vspace{-6pt}
  \caption{Performance (\underline{SE and OS})  of the models across all web domains of \underline{VisualWebArena} dataset (RQ2).}
  \label{fig:rq1_vwebarena} 
\end{minipage}
\hspace{2pt}
\begin{minipage}{0.47\textwidth}
    \centering
  \includegraphics[width=2.54in]{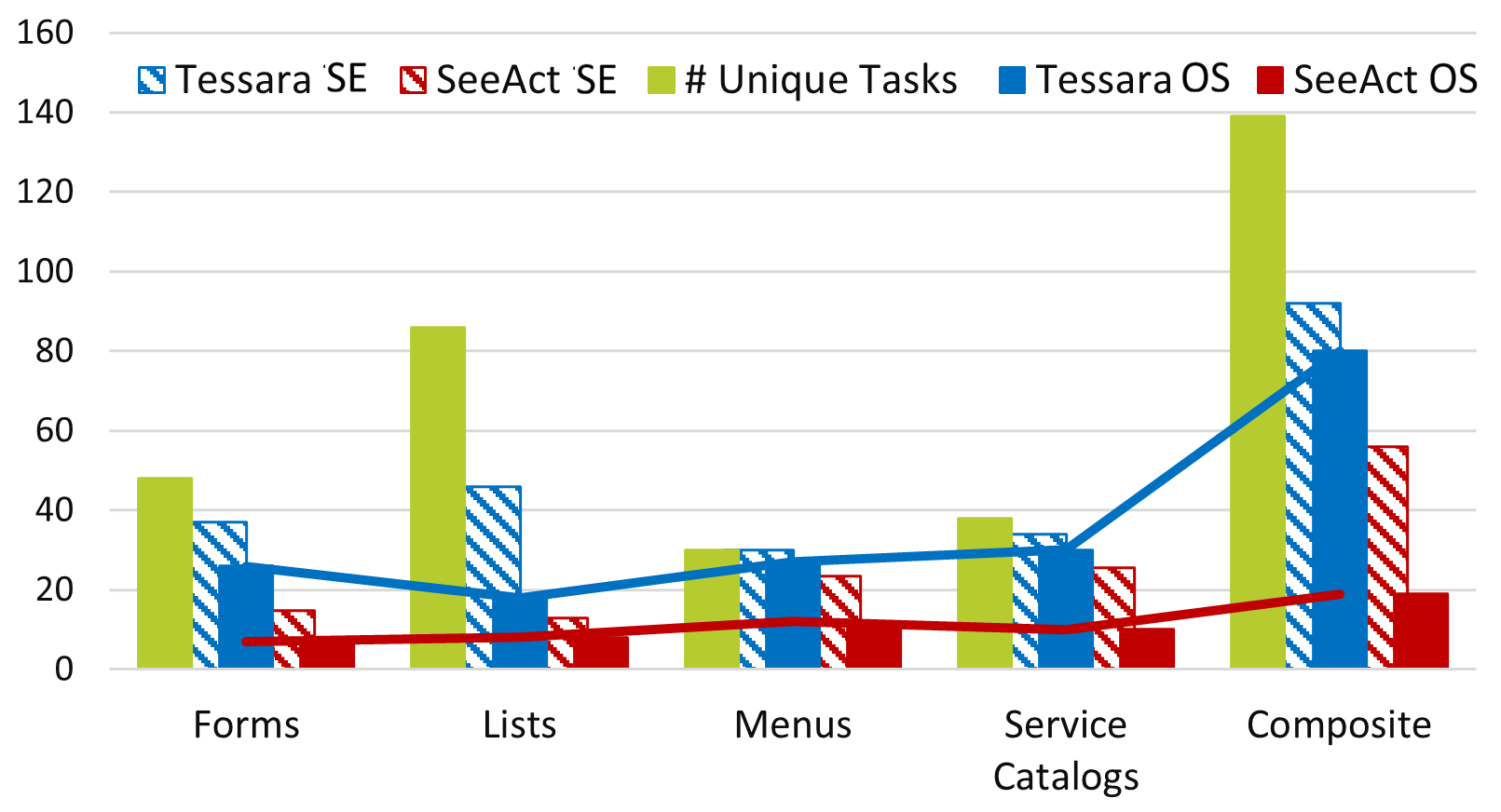}
  \vspace{-6pt}
  \caption{Performance (\underline{SE and OS}) of the models across all web domains of \underline{WorkArena} dataset (RQ2).}
  \vspace{-12pt}
  \label{fig:rq1_workarena}
\end{minipage}
\end{figure}

\begin{table*}[t]
\begin{minipage}{0.49\textwidth}
\centering
\small
\tabcolsep 3pt
\renewcommand{\arraystretch}{1.15}
\caption{\underline{Exact-/Prefix Match} of Sequence of Actions on \underline{VisualWebArena} (RQ2). Domain Acronyms: OSS:OneStopShop, Ad:advertisement, OSM: OpenStreetMap, Com.:  Composite.}
\vspace{-6pt}
\label{tab:rq1-vwebarena}
\begin{tabular}{|l|l|l|l|l|l|l|l|}
\hline
\textbf{Metrics} & \textbf{Models} & \textbf{OSS} & \textbf{Ad} & \textbf{Reddit} & \textbf{OSM} & \textbf{GitLab} &\textbf{Com.} \\ \hline
\multirow{2}{*}{\begin{tabular}[c]{@{}l@{}}Exact-\\ Match \%\end{tabular}} & SeeAct & 43.6 & 35.8 & 27.5 & 38.4 & 32 & 23.5 \\ \cline{2-8} 
 & {\tool} & 71.6 & 61.3 & 57.8 & 65.3 & 49.4 & 62.2 \\ \hline
\multirow{2}{*}{\begin{tabular}[c]{@{}l@{}}Prefix-\\ Match \%\end{tabular}} & SeeAct & 71.3 & 57.5 & 52.4 & 70.3 & 52.4 & 49.3 \\ \cline{2-8} 
 & {\tool} & 83.2 & 70.4 & 73.4 & 87.3 & 58.9 & 74.4 \\ \hline
\multirow{2}{*}{\begin{tabular}[c]{@{}l@{}}Average\\ \# Trials\end{tabular}} & SeeAct & 6 & 7 & 7 & 5 & 9 & 8 \\ \cline{2-8} 
 & {\tool} & 3 & 3 & 3 & 2 & 5 & 4 \\ \hline
\end{tabular}   
\end{minipage}
\hfill
\begin{minipage}{0.45\textwidth}
\centering
\small
\tabcolsep 3pt
\renewcommand{\arraystretch}{1.15}
\caption{\underline{Exact-/Prefix Match} of Generating Sequences of Actions on \underline{WorkArena} dataset (RQ2). Domains: Forms, Lists, Menus, S.Cat.: Service Catalog, Com.: Composite.}
\vspace{-6pt}
\label{tab:rq1-workarena}
\begin{tabular}{|l|l|l|l|l|l|l|}
\hline
\textbf{Metrics} & \textbf{Models} & \textbf{Forms} & \textbf{Lists} &\textbf{Menus} &\textbf{S.Cat.} & \textbf{Com.} \\ \hline
\multirow{2}{*}{\begin{tabular}[c]{@{}l@{}}Exact-\\ Match \%\end{tabular}} & SeeAct & 15.4 & 9.62 & 26.48 & 22.78 & 8.41 \\ \cline{2-7} 
 & {\tool} & 45.09 & 35.12 & 84.31 & 79.36 & 36.26 \\ \hline
\multirow{2}{*}{\begin{tabular}[c]{@{}l@{}}Prefix-\\ Match \%\end{tabular}} & SeeAct & 20.23 & 9.81 & 39.1 & 31.08 & 21.67 \\ \cline{2-7} 
 & {\tool} & 50.2 & 47.27 & 88.18 & 78.48 & 49.24 \\ \hline
\multirow{2}{*}{\begin{tabular}[c]{@{}l@{}}Average\\ \# Trials\end{tabular}} & SeeAct & 13 & 18 & 9 & 11 & 25 \\ \cline{2-7} 
 & {\tool} & 12 & 16 & 9 & 10 & 24 \\ \hline
\end{tabular}%
\end{minipage}
\end{table*}

\begin{table*}[t]
\caption{Effectiveness (\underline{Exact-/Prefix Match}) on Generating Sequences of Actions on \underline{WebVoyager} dataset (RQ2).}
\label{tab:rq1-webvoyager}
\vspace{-6pt}
\footnotesize
\tabcolsep 1.5pt
\renewcommand{\arraystretch}{1.15}
\begin{tabular}{|c|l|llll|ll|ll|lll|ll|ll|}
\hline
\multirow{2}{*}{Metrics} & \multicolumn{1}{c|}{\multirow{2}{*}{Models}} & \multicolumn{4}{c|}{Miscellaneous (Easy)} & \multicolumn{2}{c|}{Misc. (Hard)} & \multicolumn{2}{c|}{Repos (Easy)} & \multicolumn{3}{c|}{Repos (Hard)} & \multicolumn{2}{c|}{Google Navigation} & \multicolumn{2}{c|}{Reservations} \\ \cline{3-17} 
 & \multicolumn{1}{c|}{} & \multicolumn{1}{l|}{AllRecipes} & \multicolumn{1}{l|}{Amazon} & \multicolumn{1}{l|}{BBC-N} & \multicolumn{1}{c|}{\begin{tabular}[c]{@{}c@{}}Wolf.\\ Alpha\end{tabular}} & \multicolumn{1}{l|}{Apple} & ESPN & \multicolumn{1}{l|}{ArXiv} & \begin{tabular}[c]{@{}l@{}}Cambr.\\ Dict.\end{tabular} & \multicolumn{1}{l|}{Github} & \multicolumn{1}{c|}{\begin{tabular}[c]{@{}c@{}}Hugging\\ Face\end{tabular}} & \multicolumn{1}{c|}{\begin{tabular}[c]{@{}c@{}}Coursera\end{tabular}} & \multicolumn{1}{c|}{\begin{tabular}[c]{@{}c@{}}Google\\ Maps\end{tabular}} & \begin{tabular}[c]{@{}l@{}}Google\\ Search\end{tabular} & \multicolumn{1}{l|}{\begin{tabular}[c]{@{}l@{}}Google\\ Flights\end{tabular}} & Booking \\ \hline
\multirow{2}{*}{\begin{tabular}[c]{@{}c@{}}Exact-match \\ Accuracy (\%)\end{tabular}} & SeeAct & \multicolumn{1}{l|}{52.47} & \multicolumn{1}{l|}{\bf 50.36} & \multicolumn{1}{l|}{55.34} & 56.09 & \multicolumn{1}{l|}{44.29} & 53.34 & \multicolumn{1}{l|}{53.34} & 67.45 & \multicolumn{1}{l|}{35.32} & \multicolumn{1}{l|}{54.3} & 55.16 & \multicolumn{1}{l|}{48.2} & 60.53 & \multicolumn{1}{l|}{28.2} & 12.9 \\ \cline{2-17} 
 & \textbf{{\tool}} & \multicolumn{1}{l|}{\textbf{68.90}} & \multicolumn{1}{l|}{64.8} & \multicolumn{1}{l|}{\textbf{83.6}} & \textbf{75.33} & \multicolumn{1}{l|}{\textbf{70.78}} & \textbf{75.6} & \multicolumn{1}{l|}{\textbf{73.81}} & \textbf{76.44} & \multicolumn{1}{l|}{\textbf{52.35}} & \multicolumn{1}{l|}{\textbf{71.70}} & \textbf{65.56} & \multicolumn{1}{l|}{\textbf{66.86}} & \textbf{83.6} & \multicolumn{1}{l|}{\textbf{51.87}} & \textbf{48.67} \\ \hline
\multirow{2}{*}{\begin{tabular}[c]{@{}c@{}}Prefix Match\\ Accuracy (\%)\end{tabular}} & SeeAct & \multicolumn{1}{l|}{76.5} & \multicolumn{1}{l|}{86.59} & \multicolumn{1}{l|}{82.4} & 88.3 & \multicolumn{1}{l|}{75.5} & 83.89 & \multicolumn{1}{l|}{82.12} & 90.2 & \multicolumn{1}{l|}{61.67} & \multicolumn{1}{l|}{74.5} & 74.02 & \multicolumn{1}{l|}{83.78} & 87.4 & \multicolumn{1}{l|}{57.3} & 25.56 \\ \cline{2-17} 
 & \textbf{{\tool}} & \multicolumn{1}{l|}{\textbf{79.17}} & \multicolumn{1}{l|}{\textbf{96.24}} & \multicolumn{1}{l|}{\textbf{82.4}} & \textbf{91.4} & \multicolumn{1}{l|}{\textbf{97.2}} & \textbf{88.35} & \multicolumn{1}{l|}{\textbf{88}} & \textbf{97.8} & \multicolumn{1}{l|}{\textbf{61.39}} & \multicolumn{1}{l|}{\textbf{76.32}} & \textbf{79.74} & \multicolumn{1}{l|}{\textbf{89.3}} & \textbf{93.2} & \multicolumn{1}{l|}{\textbf{62.4}} & \textbf{67.8} \\ \hline
\multicolumn{1}{|l|}{\multirow{2}{*}{\begin{tabular}[c]{@{}l@{}}Average \\ \# of Trials\end{tabular}}} & SeeAct & \multicolumn{1}{l|}{3} & \multicolumn{1}{l|}{4} & \multicolumn{1}{l|}{5} & 3 & \multicolumn{1}{l|}{6} & 5 & \multicolumn{1}{l|}{4} & 3 & \multicolumn{1}{l|}{6} & \multicolumn{1}{l|}{5} & 4 & \multicolumn{1}{l|}{2} & 3 & \multicolumn{1}{l|}{14} & 12 \\ \cline{2-17} 
\multicolumn{1}{|l|}{} & \textbf{{\tool}} & \multicolumn{1}{l|}{\textbf{2}} & \multicolumn{1}{l|}{\textbf{3}} & \multicolumn{1}{l|}{\textbf{5}} & \textbf{2} & \multicolumn{1}{l|}{\textbf{5}} & \textbf{5} & \multicolumn{1}{l|}{\textbf{3}} & \textbf{2} & \multicolumn{1}{l|}{\textbf{4}} & \multicolumn{1}{l|}{\textbf{3}} & \textbf{4} & \multicolumn{1}{l|}{\textbf{2}} & \textbf{2}  & \multicolumn{1}{l|}{\textbf{7}} & \textbf{6} \\ \hline
\end{tabular}%
\end{table*}

\noindent {\bf Script Executability \& Oracle Satisfaction Comparison}. On WebVoyager, Fig.~\ref{fig:rq1_webvoyager} shows {\tool} and SeeAct following similar trends across domains, suggesting interface complexity affects both models similarly. However, {\tool} performs better, with the largest gap on high-complexity tasks, matching Table~\ref{tab:rq1-webvoyager}. SeeAct often performs extra actions before reaching the target page, sometimes arriving at irrelevant states.
{\tool} outperforms SeeAct across domains in VisualWebArena (see Fig.~\ref{fig:rq1_vwebarena}), with the clearest separation in Composite tasks requiring cross-site execution and state transfer. Reddit and Classifieds are challenging  due to dynamic loading, filters, and modals, yet {\tool} does better than the baseline.
On WorkArena, {\tool} improves both metrics: 66.2\% higher Oracle Satisfaction and 45.3\% Script Executability. Fig.~\ref{fig:rq1_workarena} shows improvement in every domain, especially in Composite and Service Catalog, which involve dependent workflows and hover/latency-sensitive UI patterns.


{\color{custom-blue2}\noindent {\bf Analysis on Tasks with Different Complexity Levels.} On \underline{easier, text-rich interfaces} (AllRecipes, BBC News, WolframAlpha),{\tool} generates more accurate test scripts with fewer regeneration attempts (Table~\ref{tab:rq1-webvoyager}). On dynamic websites (Apple, ESPN: (Table~\ref{tab:rq1-webvoyager})), hover menus, dynamic navigation, and changing content increase ambiguity, but step-aligned code generation and verification localize failures and selectively regenerate only the affected code fragments.}




%% file: sections/rq5.tex
\subsection{Effectiveness of Backtracking in {\model} (RQ3)}


{\textbf{Tasks where backtracking is invoked.}}
In WebVoyager, “Repo-like” domains (GitHub/arXiv/Hugging Face) exhibit the highest BT invocation rate at 55\%, exceeding “Content lookup” domains (BBC/ESPN/Wolfram) at 38\%. On content-focused websites, test script failures often stem from temporary pop-ups or being scrolled to the wrong spot. In repo-style interfaces, failures are more about crowded, complex controls (icon-only menus, nested tabs, and multi-pane layouts) which create many actions that seem correct but are actually wrong. 
The same pattern is observed in VisualWebArena: backtracking is invoked in 40\% of Classifieds/Shopping related tasks, 52\% in Reddit-like tasks, and  58\% in Composite/cross-site tasks. This gradient aligns with two sources of test fragility: 1) dynamic list mutation (faceted filters, lazy loading, infinite scroll) and 2) cross-page state transfer (composite tasks), where late-stage failures occur after long prefixes. 
Tasks in WorkArena were seen to have triggered backtracking the most: 68\% of Lists (filters/sorting) tasks invoke backtracking, whereas 59\% of Forms and 63\% for Service catalog/composite flows. 
Script execution failures often stem from subtle misapplication of filters or incorrect context rather than complete script crashes. In such cases, backtracking helps {\tool} revert to a stable script-grounded UI state and attempt a corrected sequence without full replanning.

Across WebVoyager, {\tool} invoked backtracking in 218 of 643 tasks (33.9\%). When triggered, backtracking led to successful script executability in 187 cases, corresponding to a conditional success rate of {\bf 85.8\%}. Crucially, our backtracking recovers a large number of tasks that SeeAct fails: {\em 179 of the 187  backtracked tasks (95.7\%)} were unsuccessful under SeeAct. In VisualWebArena, backtracking was used in {\bf 392 of 910} tasks (43.1\%), yielding a oracle satisfaction rate of {\bf 80.9\%}. Among these successful backtracked cases, {\em 119 tasks (30.3\%) correspond to failures under SeeAct}. WorkArena shows the highest backtracking usage, with {\bf 187 of 341} tasks (54.8\%) invoking the feature. The conditional success rate is lower than in the other benchmarks (60.4\%, 113/187), reflecting the increased complexity and longer horizons of enterprise-style workflows. Nonetheless, backtracking helps 168 tasks ({\bf 89.8\%}) that SeeAct fails to complete. This shows that a {\bf \em large number of execution failures can be resolved through backtracking rather than full replanning}.


\begin{table*}[t]
\caption{ Action types that trigger backtracking across all datasets (RQ3).}
\vspace{-6pt}
\label{tab:rq5-action-pools}
\resizebox{\textwidth}{!}{%
\begin{tabular}{|l|l|l|l|}
\hline
{\color[HTML]{000000} \textbf{Action type}} & {\color[HTML]{000000} \textbf{Share of all actions (\%)}} & {\color[HTML]{000000} \textbf{Share of BT triggers (\%)}} & {\color[HTML]{000000} \textbf{Typical reason for failure}} \\ \hline
Click \textit{(buttons/links/icons)} & 35 & 44 & \textit{Wrong button/link, mis-bound locator, collapsed menu} \\ \hline
Hover / mouseover & 8 & 15 & \textit{Panel not opened, wrong hover target, transient menus} \\ \hline
Text input \textit{(type)} & 18 & 12 & \textit{Typed in wrong field, stale placeholder, layout shift} \\ \hline
Form submit / confirm & 7 & 9 & \textit{Premature submit, wrong form instance} \\ \hline
Tab / page switch & 10 & 11 & \textit{Switched to wrong tab or sub-page} \\ \hline
Scroll / viewport move & 12 & 6 & \textit{Scrolled past target, hidden element not brought into view} \\ \hline
Other \textit{(select, checkbox, etc.)} & 10 & 3 & \textit{Incorrect option/checkbox toggles} \\ \hline
\end{tabular}%
}
\end{table*}


\begin{table*}[]
\caption{\% backtracking successes vs failures (leading to full replanning) by action category (RQ3).}
\vspace{-6pt}
\label{tab:rq5-bt-success-failure}
\resizebox{\textwidth}{!}{%
\begin{tabular}{|l|l|l|l|l|}
\hline
\textbf{Backtracked-from action category} & \textbf{BT attempts (count)} & \textbf{BT success (\%)} & \textbf{BT failure (\%)} & \textbf{Typical reasons for failure} \\ \hline
View/navigation clicks (links, tabs, menus) & 422 & 88\% & 12\% & \textit{Persistent mis-identification, ambiguous multiple links} \\ \hline
Hover-triggered menus/panels & 165 & 76\% & 24\% & \textit{Unstable hover timing, overlapping targets} \\ \hline
Pre-commit form edits (typing/select) & 221 & 82\% & 18\% & \textit{Hidden validation, auto-save changing state} \\ \hline
Scroll / viewport-only actions & 94 & 90\% & 10\% & \textit{Layout changes on re-execution} \\ \hline
Form submit / confirmation & 102 & 41\% & 59\% & \textit{Irreversible server-side commits, multi-step wizards} \\ \hline
Composite + high-level reasoning steps & 114 & 52\% & 48\% & \textit{Wrong sub-goal, misread task specification} \\ \hline
\end{tabular}%
}
\end{table*}

\textbf{Actions that Trigger Backtracking.}
Table~\ref{tab:rq5-action-pools} shows some interaction types that start the backtracking (BT). Clicks make up 35\% of all actions, but they cause 44\% of backtracking triggers. This means clicks lead to backtracking about 1.26x more often than one expects from how common they are. Most of these click failures come from clicking the wrong element (a link, icon, or an element selector that accidentally matches something hidden or collapsed). 
~Interestingly, although hover comprises only 8\% of all actions, it accounts for 15\% of BT triggers. 
This shows that hover-triggered menus are timing-sensitive, transient, and vulnerable to overlap/occlusion. 

\textbf{Action-Centric Success and Failure of Backtracking.} As seen in Table~\ref{tab:rq5-bt-success-failure}, across all categories, {\tool} attempts backtracking {\color{custom-blue2}{1,118}} times in total. Weighting by attempt counts, the implied aggregate BT recovery rate is approximately {\bf 76.7\%} (857 successes/1,118 attempts): in 3 out of 4 backtracking attempts, {\tool}’s {\bf \em backtracking converts a failing execution into a successful continuation without full replanning}.
Backtracking from scroll/viewport-only actions succeeds {\bf 90\%} of the time, and view/navigation clicks succeed 88\% of the time. These categories correspond to interactions that primarily alter client-side view state—tabs, menu expansions, navigation—where a bounded sequence of corrective actions (back, close modal, return to prior section via URL) can restore an observationally equivalent state. 

For failure cases, backtracking from form submit/confirmation succeeds only {\bf 41\%} of the time (thus 59\% failure), making this category the most difficult for BT. This is consistent with commit semantics: submission triggers server-side state transitions that cannot be undone by bounded UI operations, and even when navigation back is possible, the form instance may no longer be in the same logical state. 
The two hardest categories—submit/confirm and composite/high-level—constitute only {\color{custom-blue2}{216/1118 (19.3\%)}} of BT attempts, yet they account for {\color{custom-blue2}{60 + 55 = 115 BT failures out of 216}}, i.e., 53\% of all BT failures. 
However, this shows that despite some failed BTs for this hard task, {\em in a high percentage of tasks, {\tool} avoided full replanning with its~backtracking}.





%% file: sections/rq6.tex
\subsection{Accuracy of Test Oracle Generation (RQ4)}



This experiment aims to assess the accuracy of our three levels of verification (Section~\ref{sec:verification}). 
\underline{First}, there are 105 tasks with {\bf syntactic errors} in generated scripts. After re-prompting with compiler/runtime feedback, the scripter corrected all 105 syntactic errors.
The syntactic count is substantial since {\model} generates long, locator-heavy scripts (especially in form-based and list/table tasks), increasing opportunities for minor API/selector/import mistakes. 

\underline{Second}, for execution verification, despite the script executing successfully, {\tool} did not always land in the correct page due to the following reasons:  live web UIs are unstable (dynamic DOM, timing variance) in dynamic navigation, repo-like, and composite tasks. However, {\model} was able to reduce incorrectness to 3.5\%, resulting in {\bf 96.5\% correctness for execution~verification}.
Such failures are instances of classic web-automation brittleness rather than incorrect plans. The top two causes--element not found/detached 
and timeouts/wait conditions not met 
--account for 65\% of execution errors. These map directly to dynamic UI behaviors (lazy-loaded lists, pop-ups, re-rendered panels, slow webpages).

\underline{Third}, after successful execution of the generated script, we have the third level of semantic verification using LLM to assess if the resulting page/state aligns with the task goal. 
To assess the accuracy of the {\bf LLM semantic verifier}, we validate its reliability against manually-checked gold labels on 150 instances (chosen randomly across all 3 datasets): 90\% agreement with Cohen’s ($\kappa$=0.78), with {\bf 7\% false positives} and {\bf 3\% false negatives}. The precision of semantic verification is {\bf 81.4\%}. This asymmetry matters for test script generation: false positives should be minimized as they can certify incorrect scripts as successful. 
This confirms semantic LLM verifier as a necessary third layer beyond “no crash” execution.

%% file: sections/rq3.tex
\subsection{Ablation Study (RQ5)}
\label{sec:rq4}




\begin{table}[t]
\centering
\small
\setlength{\tabcolsep}{1pt}
\caption{Ablation Study (RQ5): Avg TS \& TC (\%)}
\label{tab:ablation_summary}
\vspace{-6pt}
\begin{tabular}{|l@{\hskip 6pt}|c@{\hskip 4pt}|c@{\hskip 4pt}|c@{\hskip 4pt}|c@{\hskip 4pt}|c@{\hskip 4pt}|c@{\hskip 4pt}|}
\hline
Variant of & \multicolumn{2}{c|}{\textbf{Shopping}} & \multicolumn{2}{c|}{\textbf{Map}} & \multicolumn{2}{c|}{\textbf{GitLab}} \\
\cline{2-7}
{\tool} & \textbf{SE\%} & \textbf{OS\%} & \textbf{SE\%} & \textbf{OS\%} & \textbf{SE\%} & \textbf{OS\%} \\
\hline
-noCoT & 67 & 40 & 63 & 42 & 53 & 35 \\ \hline
-no{\model} & 78 & 57 & 74 & 55 & 61 & 48 \\ \hline
-Verify & 82 & 65 & 79 & 63 & 65 & 52 \\ \hline
Multimodal (-noDOM) & 88 & 71 & 86 & 69 & 70 & 56 \\ \hline
Multimodal (-noPage) & 86 & 68 & 82 & 66 & 65 & 53 \\ \hline
{\bf {\tool}} & {\bf 97} & {\bf 77} & {\bf 95} & {\bf 79} & {\bf 76} & {\bf 70} \\
\hline
\end{tabular}
\end{table}

We chose Shopping, Map, and GitLab because they represent three distinct interface profiles: product search and filtering, dynamic map interaction, and dense repository-style UIs and tested on the following variants: \textbf{(i) {\tool}--noCoT}: removed CoT planning and prompted the model to generate the final script directly from the task description. \textbf{(ii) {\tool}--No{\model}}: retained CoT reasoning but omit Step-aligned Scripting ({\model}) code generation, removing alignment between reasoning and execution. \textbf{(iii) {\tool}--Verify}: disabled stepwise verification, forcing full script regeneration on failure instead of replanning from the error point. \textbf{ (iv) {\tool}--Multimodal(noDOM)}: provided only screenshots (no DOM), testing the effect of removing context. \textbf{(v) {\tool}--Multimodal(noPage)}: used only DOM (no images). 


From Table~\ref{tab:ablation_summary}, we report the following: {\em (i) removing Chain-of-Thought planning (\textbf{{\tool}--noCoT})} leads to the steepest performance drop. Without structured reasoning, the model struggles to decompose tasks and maintain goal alignment, particularly in GitLab where sequential logic and memory are critical.
{\em (ii) Disabling Step-aligned Scripting code generation (\textbf{{\tool}--no{\model}})} also substantially reduces success rates, especially in high-complexity tasks. Though reasoning exists, the lack of alignment with executable code leads to fragile execution and semantic drift.
{\em (iii) The absence of stepwise verification (\textbf{{\tool}--Verify})}  results in frequent full replanning. Both Script Executability and oracle satisfaction decrease due to repeated failures and inability to retain progress, e.g., in Map and GitLab domains.
{\em (iv) Multimodal ablations (\textbf{noDOM})} and (v) \textbf{(noPage)} show moderate drops, with DOM-only performing slightly better overall. This indicates that {\em structural information is more critical than visual context}, though both modalities contribute to robustness, especially in UI-dense websites like GitLab. 



%% file: sections/threats.tex
\section{Threats to Validity}

{\bf External Validity}: While our datasets may not be representative, they are latest benchmarks for UI agents, allowing fair comparison. Our results may vary with other LLMs. 


{\bf Internal Validity}: Regarding data leakage, the sequences of GUI interactions to finish the tasks, involving multi-step reasoning over dynamic content, are not part of static training data and are effectively unbounded. The likelihood that a model memorized the flows or interaction patterns is~low.

{\bf Construct Validity}:~Varied prompts can lead to different results. To mitigate this, we use a consistent, well-structured~prompt. LLMs can produce hallucinated reasoning/action steps. However, {\tool} has 3-tier verification. We did not compare it with the tools in the leading board for WebArena due to their unavailability.

{\bf Limitations.} {\tool} struggles with complex forms where fields are interdependent, and with dynamic pages or conditional fields that change based on input. Highly dynamic content can make automation brittle, while animated elements introduce timing issues. Captchas further block automated agents, rendering some tasks difficult. See RQ1 and RQ2 for more detailed limitations.





%% file: sections/related.tex
\section{Related Work}
\label{sec:related}

{{\bf Benchmarks} for Web Task Completion and UI Agents.}
Recent benchmarks evaluate LLM-based agents on web and GUI task completion, including WebArena~\cite{zhou2023webarena}, WebLINX~\cite{lu2024weblinx}, AssistantBench~\cite{yoran2024assistantbenchwebagentssolve}, WebGames~\cite{webgames2025}, OSUniverse~\cite{osuniverse2025}, BEARCUBS~\cite{song2025bearcubsbenchmarkcomputerusingweb}, EconWebArena~\cite{econwebarena2025}, and BrowseComp~\cite{wei2025browsecompsimplechallengingbenchmark}. 
These benchmarks report end-to-end task completion but not reusable, executable action sequences for automation-script evaluation.
Mind2Web~\cite{deng2023mind2web} includes human demonstrations, but steps are represented as natural-language actions rather than executable scripts. Thus, we use WebVoyager~\cite{he2024webvoyager}, VisualWebArena~\cite{visualwebarena2024}, and WorkArena~\cite{drouin2024workarena}, which can be augmented with canonical action sequences and replayable success conditions for our setting.

\textbf{Web, GUI, and computer-use agents.}
Agents that operate on web or GUI environments using screenshots, DOM structure, or both, include SeeClick~\cite{cheng-etal-2024-seeclick}, CogAgent~\cite{hong2024cogagentvisuallanguagemodel}, \textit{ShowUI}~\cite{lin2024showuivisionlanguageactionmodelgui}, InfiGUIAgent~\cite{liu2025infiguiagentmultimodalgeneralistgui}, UI-TARS~\cite{qin2025uitarspioneeringautomatedgui} and broader computer-use agents such as Guardian~\cite{ran2024guardian}, GPTDroid~\cite{gptdroid}, AutoDroid~\cite{autodroid}, DroidAgent~\cite{droidagent}), TaskMatrix.AI~\cite{liang2023taskmatrixaicompletingtasksconnecting}, and OSWorld/Computer Use~\cite{xie2024osworldbenchmarkingmultimodalagents}. These systems are designed for one-time task execution: they emit free-form actions, absolute coordinates, or environment-specific identifiers rather than executable, replayable browser-automation scripts. Thus, we use SeeAct~\cite{pmlr-v235-zheng24e} as a baseline, since it is the nearest prior system that can be constrained to output structured executable actions without adding a custom post-processing layer.

\textbf{Code-grounded reasoning.} {\tool} is related to CoT~\cite{wei2023chainofthoughtpromptingelicitsreasoning} and Program-of-Thought (PoT)~\cite{chen2023program}, which use explicit intermediate reasoning and/or executable code. Existing PoT studies focus on structured reasoning tasks such as math, logic, and code  generation~\cite{bi2023programofthoughtsworkreasoning}, rather than web testing.

%% file: sections/conclusion.tex
\section{Conclusion}
\label{sec:limitations}


{\tool} advances LLM-based web automation test generation by addressing adaptability and step-level failure localization. It integrates CoT-based planning with Step-aligned Scripting ({\model}), which turns each predicted action into executable code for stepwise verification, backtracking, and targeted recovery. By combining multimodal inputs from the DOM and screenshots, {\tool} enables adaptive re-planning when execution fails.
Overall, {\model} demonstrates the value of stepwise, verifiable reasoning for reliable web automation agents. This opens future directions for incorporating richer inputs, such as video, voice, and logs, and supporting deeper UI interactions in automated web testing.